\documentstyle[12pt]{article} 
\newcommand{\be}{\begin{equation}}
\newcommand{\ee}{\end{equation}}
\newcommand{\beq}{\begin{eqnarray}}
\newcommand{\eeq}{\end{eqnarray}}
\newcommand{\bear}{\begin{array}}
\newcommand{\ear}{\end{array}}

\begin{document}
\title{Starobinsky Model in Schroedinger Description}
\author{S.Biswas $^{*a),b)}$, A.Shaw $^{**a)}$ and D.Biswas$^{a)}$ \\
a) Department of Physics, University of Kalyani, West Bengal,\\
 India, Pin.- 741235 \\
b) IUCAA, Post bag 4, Ganeshkhind, Pune 411 007, India\\
$*$ email: sbiswas@klyuniv.ernet.in\\
$**$ email:amita@klyuniv.ernet.in}
\date{}
\maketitle
\qquad\qquad\qquad decoherence, wavefunction of the universe
\begin{abstract}
In the Starobinsky inflationary model inflation is driven by quantum 
corrections to the vacuum Einstein equation. We reduce the Wheeler-DeWitt 
equation corresponding to the Starobinsky model to a Schroedinger form 
containing time. The 
Schroedinger equation is solved with a Gaussian ansatz. Using the prescription 
for the normalization constant of the wavefunction given in our previous
work, we show that the Gaussian ansatz demands Hawking type initial conditions for 
the wavefunction of the universe. The wormholes induce randomness in initial 
states suggesting a basis for time-contained description of the Wheeler-DeWitt
equation.                  
\end{abstract}
\section{\bf{Introduction}}
Inflation is an essential ingredient in modern cosmology to solve the
horizon, the flatness and the monopole problems. In order to ascertain 
whether inflation is natural, there have been several attempts to 
study the curvature squared cosmology along with the Einstein curvature 
term. Even before Guth's [1] proposal of inflationary scenario, 
Starobinsky [2] proposed an inflationary model taking one loop 
quantum corrections to the classical Einstein Lagrangian. It has been 
shown that this model is equivalent to the curvature squared model in 
which the universe tunnels into the de Sitter phase that becomes unstable 
and finally emerges into the Friedmann era. 
\par         
In order to show that the inflation is quite natural 
phenomena for any initial conditions (called no hair 
conjecture), several authors [3] attempted to show that, for some 
models, inflation is really an attractor. The works done by 
Mijic, Moris and Suen [4], Maeda [5] and Vilenkin [6-7] refer 
to the early universe situation of inflationary cosmology. Vilenkin 
deals with a quantum viewpoint to study the Starobinsky model and 
considers the tunneling phenomenon in the de Sitter phase through 
solution of the Wheeler-DeWitt equation [7] in quantum 
cosmology. In this approach one requires a specific ansatz for the 
initial conditions, that are currently divided into two ways: the Vilenkin 
proposal [8] and the Hartle-Hawking proposal [9]. In the 
approach (without phase transition), the de Sitter solution is obtained 
as self consistent 
solution of the vacuum Einstein equations modified by one loop 
quantum corrections due to quantized conformal matter fields or 
through the solution of the Wheeler-DeWitt equation with 
appropriate boundary conditions leading to an inflationary 
de Sitter modes. The other approach is related to phase 
transitions. Here the inflation is driven by a false vacuum energy 
density and is related to the inflaton field in the model. The no hair 
conjecture is found to exist in both type of approaches as 
revealed by the works of Starobinsky and Schmidt [10]. In Maeda [5] 
as well as in Mijic et al [4], the analysis of the inflation is looked 
upon in a classical way. Using Wald's [11] strong and dominant 
energy condition, it has been shown that in $R^2$-type cosmology 
inflation is quite natural but acts as a transient attractor 
only. In Mijic, Morris and Suen the curvature squared theory has 
been shown to be equivalent to the Einstein gravity theory with a 
scalar field but without any potential form for the scalar 
field. Though Maeda succeeds in obtaining a potential form in 
their work with a long plateau region, it is not  
clear how the universe from a region of Planck region arrives at 
the plateau region simply by losing energy  
to roll over the flat region depending on initial 
conditions. We believe, the answer should come through the quantum 
behaviour, which is absent in their model [4,5]. The quantum analysis 
carried out by Vilenkin [7] looks nice. However the initial 
condition of tunneling behaviour with respect to no hair 
conjecture has not been tackled with justice. However Starobinsky 
analyzed his $R^2$ type theory to show that there is an unstable 
de Sitter solution followed by the present Friedman era after 
sufficient inflation. Though Vilenkin tried to answer some aspects 
of stability of de Sitter solution in the Starobinsky model, but it is 
not clear even now how does the self consistent de Sitter solution 
in the Starobinsky model exits to the Friedmann era, though  
it is unstable in both past as well as in the future. We try to understand this
aspect from the energy conservation. 
\par
Moreover the curvature fluctuation term has been introduced in an 
adhoc way. In calculating the curvature fluctuations, Vilenkin treats 
the wavefunction as if it is normalized but it is known that the normalization 
of Vilenkin 
wavefunction is awkward. Whereas the instability is related to the curvature 
fluctuations and which in turn is related to the randomness in the initial states it is
therefore necessary to study the origin of randomness. As the Starobinsky solution
emerges due to the quantum corrections, the quantum superposition principle remains
inherent in this description. Hence the emergence of the classical de Sitter 
universe should be answered in a natural way from the study of the quantum 
wavefunction of the universe. 
\par
In this paper we try to answer this problem at least 
being successful to point out the direction in which the 
investigation has to be persued to get an final answer. Our aim is to
study (i) the normalization aspect of the wavefunction, (ii) the
tunneling probability and (iii) the randomness and the curvature
fluctuation in the Starobinsky description. It is 
shown that the energy momentum conservation forces a de Sitter 
universe to tunnel into a realistic spacetime. The tunneling from 
the Euclidean to the Lorentzian sector also demands a Wheeler-DeWitt type 
equation for the quantum wavefunction (referred as instanton) 
to settle the initial conditions which in turn is related to the 
two proposals of the quantum wavefunction of the universe. Here 
we find that the wormhole connected universe greatly modifies the 
initial conditions to be imposed on the Euclidean de Sitter 
instanton universe. We report in this paper the main results.
\par         
The quantum instability of the de Sitter spacetime has been 
discussed at length by Mottola [12] and many others to 
understand the nature of $Im\; L_{eff}$ with respect to the de Sitter 
spacetime. It may be pointed out that the $Im\; L_{eff}$ is a measure of 
the instability. 
From these works and from the work of Nevelli [13], it 
is clear that the tunneling paths in the Euclidean sector are crucial to 
understand the instability of the de Sitter vacuum. But there are no 
successful attempts to include such a contribution  
in the Starobinsky 
type model. In section II we discuss the Starobinsky model with 
some known results, barely needed in our follow up discussion 
and also discuss the tunneling of the universe in the de Sitter 
space, fixed by the energy momentum conservation. This treatment 
allows us to persue the quantum evolution of the universe in the 
context of the Wheeler-DeWitt equation and study the evolution of the 
universe in $R^2$ cosmology. In section III we introduce a time variable 
to reduce the Wheeler-DeWitt equation to the Schroedinger form
(containing time) of quantum 
mechanics. The section IV deals with the calculations of the wavefunction in 
the Starobinsky model. The section V ends with a discussion. 
\section{\bf{The Starobinsky model}}
In the Starobinsky model inflation is driven by the quantum 
corrections to the vacuum Einstein equations 
\be
{R_{\mu \nu}} - {1 \over 2} {g_{\mu \nu}} R = - 8 \pi G < {T_{\mu \nu}} >,
\ee
where $ < T_{\mu \nu} >$ is the quantum corrections and is given by
\be
< {T_{\mu \nu}} > = { \alpha \over 6}^{\;\;(1)} 
{H_{\mu \nu}}+{\beta^{\;\; (3)}}{H_{\mu \nu}}, 
\ee
where
\be
{(1)}_{_{H_{_{\mu \nu}}}}={R_{\,;\,\mu \, ; \, \nu}}-{g_{\mu \nu}}
{{R_{\, ; \, \sigma}}^{\, ; \, \sigma}}+R{R_{\mu \nu}}
-{1 \over 4} {g_{\mu \nu}}R^2
\ee
and
\be
{{{(3)}_{_{H}}}_{_{\mu \nu}}}={R_{\mu}^{\sigma}}{R_{\nu \sigma}} -{2 \over 3}
R{R_{\mu \nu}} -{1 \over 2}{g_{\mu \nu}}{R^{\sigma \tau}}{R_{\sigma \tau}}
+{1 \over 4}{g_{\mu \nu}}R^2.
\ee
We use the following conventions
\be
{R^{\alpha}_{\beta \gamma \delta}}={{\partial}_{\delta}}{\Gamma^{\alpha}
_{\beta \gamma}}-{\partial}_{\gamma}{{{\Gamma}^{\alpha}}_{\beta \delta}}
-{{{\Gamma}^{\alpha}}_{\mu \gamma}}{{{\Gamma}^{\mu}}_{\beta \delta}}
+{{{\Gamma}^{\alpha}}}_{\mu \delta}
{{{\Gamma}^{\mu}}_{\beta \gamma}}.
\ee
\be
{R_{\mu \nu}}={{R^{\alpha}}_{\mu \alpha \nu}}.
\ee
and the Ricci identity as 
\be
{{A^{\alpha}}_{\, ; \, \mu \, ; \, \nu}}-
{{A^{\alpha}}_{\, ; \, \nu \, ; \, \mu}}={A^{\beta}}{{R^{\alpha}}_
{\beta \mu \nu}},
\ee
with ${{R^{\alpha}}_{\beta \gamma \delta}}$ antisymmetric in last two indices. 
The metric is assumed to be of the Robertson-Walker form,
\be
d{s^2}={C(\eta)}{\left[ {d{\eta}^2}-{d {{\Sigma}_{k}^{2}}(r, \theta, \phi)}
\right ]},
\ee
where $d{{{\Sigma}_{k}}^{2}}$ is the metric on a 3-sphere, 3-plane and 
3-hyperboloid 
for closed $(k = +1)$ , flat $(k=0)$ and open $(k=-1)$ metric 
respectively. As we would need some properties of ${{{(1)}_H}_{\mu \nu}}$ 
and 
${{{(3)}_H}_{\mu \nu}}$ 
in our description, we mention here some salient features. 
The tensor ${{{(1)}_H}_{_{\mu \nu}}}$ is identically conserved. Using the Bianchi
identity 
\be
{({R_{\mu \nu}} - {1 \over 2} {g_{\mu \nu}} R)}_{\, ; \, \nu} = 0,
\ee
we get from (3)
\be
{{(1)}_H}_{\mu\,; \nu}^\nu={{R^{ \, ; \,\nu}}_{\, ; \, \mu \, ; \, \nu}}
-{g_{\mu}^{ \nu}}
{{R_{\, ; \, \sigma \, ; \,  \nu}}^{ \, ; \,\sigma}}
+{R_{ \, ; \,\nu}}R^{\nu}_{\mu}.
\ee
Defining ${R^{; \nu}} = A^{\nu}$ and using Ricci identity Eqn.(7), we get
\be
{{(1)}_{{H^{\nu}}_{\mu ; \nu}}}=0.
\ee
The tensor  
${{{(3)}_H}_{\mu \nu}}$,
on the other hand is conserved only in the
conformally flat spacetime and can be written as
\be
{{{(3)}_H}_{\mu \nu}}={1 \over 12}{R^2}{g_{\mu \nu}}-{R^{\rho \sigma}}
{R_{\rho \mu \sigma \nu}} .
\ee
The trace of the above two tensors is given by 
\be
{{{(1)}_H}_{\mu}^{\mu}}=-3{{R_{; \nu}}^{;\nu}}
\ee
and
\be
{{{(3)}_H}_{{\mu}}^{ \mu}}={1 \over 3} R^2-{R_{\alpha \beta}}{R^{\alpha \beta}}.
\ee
The sum of the Eqns.(13) and (14) gives the so called trace anomaly 
i.e., the regularization carried out in n-dimension keeps a 
footprint when put back in 4-dimension and thus breaks the 
conformal invariance of the original action. In the RW spacetime
since we have a single variable $C(\eta)$, it is sufficient to 
consider the time-time components of Eq.(1); all other components 
will be linearly related. We write 
${{(1)_{_{H}}}_{_{00}}}$ and
${{(3)_{_{H}}}_{_{00}}}$ explicitly.
\be
{{{(1)}_{_{H}}}_{_{00}}}={1 \over 2}C^{-1}[ -9 { \ddot{D} }D +{9 \over 2}{{ \dot{D} }^2}
+{27 \over 8}{D^4}+9k{D^2}-18{k^2}],
\ee
\be
{{{(3)}_{_{H}}}_{_{00}}}={C^{-1}}{[{3 \over 16}{D^4}+{3 \over 2}k{D^2}+3{k^2}]},
\ee
\be
{R_{00}}-{1 \over 2}{g_{00}}R = -3{({{D^2} \over 4}+k)}.
\ee
Here $D= {{ \dot{C} }\over C}$ and the dot refers to the derivative with respect to 
$\eta$. Introducing
\be
d \eta={dt \over a(t)},
\ee
we get the `00' component of Eqn.(1) as
\be
{{{{ a'}^2}+k} \over {a^2}}= {1 \over {H^{2}_{0}}}{\left [
{{{{ a'}^2}+k} \over {a^2}}\right ]}^{2}- {1 \over {M_{0}}^2}{\left[{
{2{ a'}{ a'''}} \over {a^2}}
-{{{ a''}^2} \over {a^2}}
+{{2{{ a''}}{ a'}^2} \over {a^3}}
-{3 ({{ a'} \over a})^4} -2k {{{ a'}^2} \over a^2} + {k^2 \over a^4}
\right]}.
\ee
where all constants $\beta , \alpha $ and $G$  are incorporated in 
$H_0$ and $M_0$. The 
Eqn.(19) has solutions
\be
a(t)={H_0}^{-1} {\cosh {{H_0}t}},\qquad k=+1,
\ee   
\be
a(t)={a_0} {\exp {({H_0}t})},\qquad \qquad k=0,
\ee   
\be
a(t)={H_0}^{-1} {\sinh {{H_0}t}},\qquad k=-1,
\ee   
Eqns.(1) and (19) refer to the Starobinsky description. In quantum 
language it is said that the universe tunnels quantum 
mechanically from ``nothing'' to the de Sitter  
inflationary mode, (Eqns.(20,21,22)). Here ``nothing'' means no classical spacetime. 
At the moment of nucleation, $t= 0$ the universe has a size 
$a(0)=H^{-1}$, zero velocity i.e., $a'(0)=0$. This is the beginning of 
time. In quantum theory, tunneling is mainly due to vacuum 
fluctuations. The quantum cosmology deals with the tunneling of 
universe from the Euclidean region $a < H^{-1}$ usually termed as 
the classically forbidden region to a classically allowed region 
$a>H^{-1}$. There are several modes of inflationary models, among which 
the de Sitter mode of evolution plays a key role. If the de 
Sitter phase plays any previledged role with signature change 
occurring at the moment of nucleation, one has to know the 
behaviour of quantum wavefunction of the universe in the 
Lorentzian regime from a path integral approach or from the tunneling 
approach. 
\par        
To decide some of this aspects and also the stability of the
Starobinsky phase we note the following exercise. Let us write 
${{{(1)}_{_H}}_{\mu \nu}}$ 
term in two parts 
\beq
{{(1)}_{_{H}}}_{_{\mu \nu} }=[{{R_{\, ; \, \mu \, ; \, \nu}-{g_{\mu \nu}}
{{R_{\, ; \, \alpha}}^{\, ; \, \alpha}} - RR_{\mu \nu} 
+ {{g_{\mu \nu} R^2} \over 4}}]
+[2R( {R_{\mu \nu}} - {{g_{\mu \nu} R} \over 4})}] \nonumber \\
={(1)_{{\tilde{H}}_{\mu \nu}}}+{(1)_{S_{\mu \nu}}}
\eeq
and neglect the second term to construct $< {T_{\mu \nu}} >$.
The $< {T_{\mu \nu}} >$ so
constructed is not conserved as is evident from Eqn.(10) and 
(11). Neglecting  
${(1)_{S_{\mu \nu}}}$ term we get
\be
2^{\;\;(1)}{{H}_{00}}={1 \over C}{(\, -9D \ddot{D} + 
{45 D^4 \over 8}-{9 {\dot{D}^2} \over 2}+27 k {D^2}C +18 k^2)}.
\ee
Using Eqn.(24) and (16) in Eqn.(1) we get for the `00' component of 
the Einstein equation as 
\be
{{{{ a'}^2}+k} \over {a^2}}= {1 \over {H^{2}_{0}}}{\left [
{{{{ a'}^2}+k} \over {a^2}}\right ]}^{2}- {1 \over {M_{0}}^2}{\left[{
{2{ a'}{ a'''}} \over {a^2}}
+{{{ a''}^2} \over {a^2}}
+{{2{ a''}{ a'}^2} \over {a^3}}
-{{5 {a'}^4} \over a^4}  -6k{{{a'}^2} \over a^4}- {k^2 \over a^4}
\right ]},
\ee
with $a'={{\partial a} \over {\partial t}}, d \eta = {{dt} \over {a(t)}}$. 
All constants $\beta, \alpha $ and $G$ are 
incorporated in $H_0$ and $M_0$. The Eqn.(25) has a characteristic 
feature that, it has also the same set of solutions (20-22) as well 
as the Euclidean de Sitter solution for $k =+1, \,0, \,-1$ but  
${{< T_\mu ^\nu \,> \,} _{; \, \nu}} \neq 0$,  as is evident from the 
expression of ${{(1)}_S}_{\mu \nu}$. 
To show it we convert 
Eqn.(25) setting $H = {a'\over a}$ as
\be
H^2 {(H^2 - {H_0}^2 )} = {{H_0}^2 \over {{M_0}^2}}{(2H\, H'' +10 H^2 H' 
+ {H'}^2)}.
\ee
This shows that $H = H_0$ is a solution of Eqn. (26). This result 
seems quite interesting. In realistic spacetime 
we know ${T^{\nu}_{\mu \, ; \, \nu}} =  0$. So 
emergence in the de Sitter mode is possible only if  
${(1)_{S_{\mu \nu}}} =0 $. From
the expression given in Eqn.(23), we find that for the de Sitter type 
of solution  
${(1)_{S_{\mu \nu}}} $ exactly vanishes.
This explains the emergence of de Sitter 
phase in the Starobinsky description. As soon as the universe evolves 
into the Starobinsky type de Sitter phase we have $t = 0$ i.e., the 
universe starts at $a = H^{-1}$. As we observe the de Sitter mode of 
solution even with  
${{< T_\mu ^\nu \,> \,} _{; \, \nu}} \neq 0$, 
the existence of such a solution is 
permitted provided the violation obeys 
\be
{(\Delta T_{00})} ( \Delta \tau) \, >> 1,
\ee
where $\Delta \tau \, << the\;Planck\; Time.$ The Eqn.(27) then implies  
$\Delta T_{00} \, >> 10^{19} GeV$
i.e., we have a hot big bang scenario with temperature   
$\approx 10^{17} GeV.$ 
\par
It is quite probable that at such a short scale 
$(10^{-33} cm)$ higher order corrections like $R^3, R^4$ etc. begin to 
compensate the violation. However, whatever be the 
magnitude of compensation, the violation is there and the only 
way the universe tunnels to a realistic spacetime is through the 
de Sitter mode of evolution making  
${(1)_{S_{\mu \nu}}} =0 $
and there is a 
signature change at the moment of nucleation. The universe 
tunnels from the Euclidean de Sitter to the Lorentzian de Sitter 
solution. Using the expression for ${(1)_S}_{_{\mu\nu}}$, we get 
\be
\vert \, \Delta T_{00} \, \vert \approx {{k^2} \over C} \approx {{k^2} \over
a^2(0)},
\ee
taking $a'(0)= 0$ around the nucleation point. Now $a(0)= H^{-1}$, so 
\be
\vert \, \Delta T_{00} \, \vert \approx {{k^2}{H^2}} . 
\ee
At this point one may be tempted to argue that the tunneling to the
Starobinsky phase occurs with higher probability when $H$ gets 
smaller. But we then have to abandon the inflationary scenario 
and its fruits related to the horizon and the flatness problem. However 
there is a dynamical zeroing of $\vert \Delta T_{00} \vert $ 
leading to de Sitter mode 
causing
${(1)_{S_{\mu \nu}}}=0$.
In order to incorporate this dynamical aspect 
in the framework of the quantum cosmology we lean towards the 
calculation of tunneling probability to the Starobinsky phase. 
The treatment will help us answer about the nature of the boundary 
conditions that would satisfy the solution of the Wheeler-DeWitt equation.
\par
It should be pointed out that while subtracting a part from the energy
momentum tensor, as if we spoiled the energy momentum conservation still 
getting the de Sitter solution with the truncated $T_{oo}$. What we 
observe that only for the de Sitter solution, $(1)_{_{{\sc{S}}_{\mu\nu}}}$
term is automatically zero and the energy conservation is restored. For any 
other solution energy momentum conservation would be violated. This is the 
reason of spontaneous nucleation of the Starobinsky phase into the de Sitter 
phase, hitherto not pointed out in any previous work. One should not confuse 
that we get de Sitter solution even violating the energy conservation. This 
would, if true, then leads to horizon size fluctuations at the onset of the 
FRW hot big bang destroying the benefits of inflation. What we like to point 
out that the uncertainty principle plays a decisive role in setting the 
quantum character of the universe as well as the initial conditions.  
\section{\bf{Time in Quantum Gravity and 
The Schroedinger Equation}}
The discussion in section II reveals that the quantum corrections to the vacuum 
Einstein's equation drive the inflation in the Starobinsky model and there is a
transition from the Euclidean to the Lorentzian region. Moreover, the transition to
the classical spacetime occurs for $ {T_{\mu}^{\nu}}_{\, ; \nu} = 0$, with a
de Sitter inflation. There occurs thus a quantum to classical transition in which
energy conservation plays a deciding role as if it acts as a boundary condition 
for the transition. 
\par
In the classical description, the time and the Hamiltonian are related through
the Hamilton's
equation whereas the Schroedinger equation plays the same role in the quantum 
description. But in quantum gravity, because of the constraint relation 
$H = 0$, quantization of the gravitational field faces a conceptual problem in
that no time parameter appears at a fundamental level. This is the problem of 
time in quantum gravity and recently the matter is under serious investigation 
by various authors [14-20]. 
\par
The energy conservation, the Schroedinger equation (the Wheeler-DeWitt equation in
quantum gravity) and the interpretational framework require the introduction of 
an external time parameter to understand in a coherent way the quantum to classical
transition and the tunneling problem in the quantum gravitational description. We
will be mainly concerned with these two aspects in the Starobinsky description. The
emergence of universe in classical region requires the solution of the
Wheeler-DeWitt equation and the effectiveness i.e., the interpretational framework
will be discussed in studying the decoherence mechanism in quantum gravity. 
\par
It has been shown by Santamato [21] how to arrive at 
the Schroedinger equation starting 
from classical description using Madelung-Bohm [22,23] and Feynes-Nelson [24,25] approaches
bearing epistemological content of traditional quantum mechanics. He starts
with a function $S(q,t)$ satisfying the Hamilton-Jacobi equation of 
the classical mechanics
\be
{{\partial S} \over {\partial t}} + H(q, {\nabla S}, t) = 0
\ee
where 
$H(q, {p = \nabla S}, t)$ is the classical Hamiltonian such that 
\be
{v^i} (q,t) = {{\partial H{(q, \nabla S, t )}}\over {\partial p_i}}
\ee
and following Madelung-Bohm and Feynes-Nelson approaches defines a probability 
density $\bar{\rho}(q,t)$ satisfying the continuity equation
\be
{\partial_{_{t}}}\; \rho + {\partial_i} ({\bar{\rho}} v^i) = 0.
\ee
It has been then shown that the function 
\be
\psi (q,t) = {\sqrt{\rho{(q,t)}}} \exp {\left[{i \over \hbar} S (q,t)\right]}
\ee
satisfies the Schroedinger equation for a given Hamiltonian combining (30) and (32)
with gradient of scalar curvature of space (the Weyl space) introducing randomness
on initial positions. 
\par
As there is no time variable in quantum gravity we demand $S (q,t) = S(q)$ and
it satisfies the Hamilton Jacobi equation but source free. Defining a time parameter,
we call it WKB time, it is obvious that 
${\sqrt{\rho{(q,t)}}} \propto \psi (q,t)$ 
then satisfies the Schroedinger equation when $S(q,t)$ is time independent.
Though not identical, the same view is also expressed by Kiefer [18,19], in defining
a time operator $d \over dt$ starting from a minisuperspace Wheeler-DeWitt equation.
We discuss briefly the outline of our approach with minimally coupled scalar 
field in gravitational background. The details is placed elsewhere
[26]. In the next 
section the approach will be elucidated for the Starobinsky description. 
\par
We start with an action
\beq
I & = & {\int {{d^4} x {\sqrt{-g}}\left[ {-R \over {16 \pi G}} - 
{1 \over {2 {\pi}^2}} ({1 \over 2} \phi_{,\, \mu} \phi^{,\,\mu} 
+ V(\phi) ) \right] } } \nonumber \\
& - & {1 \over {8 \pi G}}{\int_{\sum} d^3 x \sqrt{h} K}
\eeq
in the FRW model
\be
ds^2 = - dt^2 + a^2 (t) \left[ {{{dr^2} \over {1 - k r^2}} + r^2 (d {\theta}^2
+{\sin}^2 \theta d {\phi}^2)} \right].
\ee
The Hamiltonian constraint is
\be
-{1 \over {2Ma}} {P_a}^2 + {1 \over {2 a^3}}{P_{\phi}}^2 - {M \over 2} ka 
+ a^3 V(\phi) = 0.
\ee
The dynamical equations are
\be
{{\ddot{a}} \over a} = - \left[ {{{\dot{a}}^2} \over {2a^2}} + {k \over {2a^2}}
+ {3 \over M}\{{1 \over 2}{\dot{\phi}}^2 - V(\phi)\} \right]
\ee
and
\be
{\ddot{\phi}} + 3 {{\dot{a}} \over a} {\dot{\phi}} 
+ {{\partial V} \over {\partial \phi}} = 0\,.
\ee
In (34) to (37),
$M =  {{3 \pi}\over {2G}} 
   = {{3 \pi {m_{Planck}^{2}}} \over 2}, 
k  =  0, \pm 1 $ and
\be
P_a = - M a {\dot{a}},\;
P_{\phi} = a^3 {\dot{\phi}}.  
\ee
The prime denotes derivative with respect to time and $k$ is the trace of extrinsic 
curvature. Identifying 
$P_a = {{\partial S_o} \over {\partial a}},
P_\phi = {{\partial S_o} \over {\partial \phi}} $,
the Hamilton-Jacobi equation is
\be
{-{ 1 \over {2M}} }{({{\partial S_o} \over {\partial a}})^2} + {1 \over {2a^2}}
{({{\partial S_o} \over {\partial \phi}})^2} -{1 \over 2}Mka^2 + a^4 V(\phi) = 0.
\ee
We define a time operator (a directional derivative)
\be
{\partial \over {\partial t}} = \sum_{i} ({{{\partial H} \over {\partial P_i}}
{{\partial} \over {\partial q_i}} - {{\partial H} \over {\partial q_i}}
{{\partial} \over {\partial P_i}}}).
\ee
Going to the FRW minisuperspace, we get the Wheeler-DeWitt equation
$H \psi = 0$ i.e.,
\be
\left[ {{\hbar^2} \over {2M}} ({\partial^2 \over {\partial a^2}} 
+ {p \over a}{{\partial} \over {\partial a}})
+ {\hbar^2 \over {2a^2}} {\partial^2 \over {\partial {\phi^2}}} 
- {{Mka^2} \over 2} + a^4 V(\phi) \right] \psi = 0.
\ee
Identifying 
$P_a = - i \hbar {\partial \over {\partial a}}$
and
$P_{\phi} = - i \hbar {\partial \over {\partial \phi}}$ and
substituting
\be
\psi (a, \phi) = \exp \left[ i {\sum_{n=0}^{\infty}} A_n M^{1-n} \right]
\Phi(a,\phi)
\ee
in (42) we find in ${M^2}- order \; \, {{\partial A_o} \over {\partial \phi}} = 0$,
and in ${M^1}-order$
\be
{a^2 \over 2}({{\partial A_o} \over {\partial a}})^2 + {1 \over 2} k a^4 = 0.
\ee
Identifying $S_o = MA_o$, equation (44) gives the source free Hamilton-Jacobi
equation see (40):
\be
{a^2 \over {2M}} {P_a}^2 + {M \over 2}ka^4 = 0.
\ee
Using this condition in (41) we get
\be
{d \over {dt}} = - {1 \over {Ma}}{{\partial S_o} \over {\partial a}}{{\partial}
\over {\partial a}}, 
\ee
since $S_o = S_o (a)$ only. Thus 
writing
\be
\psi (a, \phi) = {e^{i A_o (a) M}}\;\; \Phi (a, \phi)
\ee
and substituting (47) in (42) and using (44) and (46) we get
\be
i \hbar {{\partial \Phi} \over {\partial t}} \simeq H_{\phi} (a, \phi) \;\;\Phi
\ee
neglecting ${\hbar^2}-order$ term ${{\partial^2 \Phi} \over {\partial a^2}}$.
As mentioned in the beginning of this section, we obtain the reduction (48) using
(41), instead of the continuity equation. In the next section we take up the Starobinsky
model to evaluate ${\vert \psi \vert}^2$ using a Gaussian ansatz for 
$\Phi (a, \phi)$. 
\par
One might point out various drawbacks in obtaining (48)
using (41), (43) and (46), which are now been persued by many authors in
the framework: canonical quantization of gravity [27,28]. Recently we
have been able to obtain (48) using the prescription of `time before
quantization' [26] . One may consider (48) from a different angle.
Suppose we have (48) valid in the semiclassical region, we construct
$H_\phi$ for the Starobinsky description, adopt a boundary condition for
$\phi$ (of course valid in the large scale factor region), investigate
whether the boundary condition is anyhow related to the boundary
condition at the small scale factor region and reproduces the
wavefunction of the universe suiting a given boundary condition
proposal.
\section{\bf{Wavefunctions in the Starobinsky model.}}
The main problem in effecting canonical quantization in the Starobinsky description
in minisuperspace formalism is the absence of an action in closed form. However it 
has been shown that [7] for ${M_o}^2 << {H_o}^2$ an action in closed form is 
obtained as
\be
S = 2 {\pi^2}{\int L (R) a^3 dt}
\ee
with
\be
L (R) = {1 \over {16 \pi G}}( R + {{R^2} \over {6 {M_o}^2}} 
+ {{R^2} \over {R_o}} \ln {R \over R_o})
\ee
where $R_o = 12 {H_o}^2$. Evaluating the action it is found that it has maximum at
$R = R_o$ so that the semiclassical tunneling probability is given by
\be
P \propto \exp {(- \vert S_o \vert)} = \exp {(- {{4 \pi} \over {GM_o^2}})}.
\ee
The curvature fluctuation is obtained by expanding $S(R)$ around $R = R_o$
\be
S(R) = S_o + {1 \over 2}S'' (R_o)(\delta R)^2.
\ee
Using (49) and (50), $S(R)$ is given by
\be
S(R) = {{24 \pi } \over G}({1 \over R} + {1 \over {6 {M_o}^2}} 
+ {1 \over R_o^2} \ln {R \over R_o})
\ee
so that the curvature fluctuation
\be
({{\delta R} \over R_o})^2 \sim {G {H_o}^2 \over \pi}.
\ee
\par
It is interesting to note that this curvature fluctuation is proportional to
$\vert \Delta T_{oo} \vert$ as is found in (28). However the result (54) is
classical. An informative description for the nucleation of the universe can 
be obtained by solving the Wheeler-DeWitt equation  for the wavefunction of 
the universe. Vilenkin [] attempted this problem solving (42) for the Starobinsky 
model with the tunneling
boundary condition to obtain a curvature fluctuation of the order of (54).
Our aim is to study these aspects with the formalism just mentioned above to
explore:\\
 (i). the normalization aspect of the wavefunction\\ 
(ii). the tunneling probability \\
(iii). the randomness and the curvature fluctuation\\ 
(iv). the randomness and the decoherence and \\
(v). the backreaction and the wormholes contribution to the solution of 
the Wheeler-DeWitt equation. In this paper we take up the first three points.
\par
Using the method outlined in [7], the Wheeler-DeWitt equation reads 
\be
\left[ {\partial^2 \over {\partial q^2}} - {1 \over q^2}{\partial^2
\over {\partial x^2}} - V(q,x) \right] \Psi (q,x) = 0
\ee
where $q = H_o \,a \,({L' \over {{L'}_o}})^{1 \over 2}$ with $L' = L'(R)$ and
${L_o}' = L'(R_o)$, and
\beq
 x & = & {1 \over 2} \ln ({R \over R_o}),  \\
V(q,x) & = & \lambda^{-2} q^2 (1 - q^2 + \mu^2 (x) q^2),  \\
\mu^2(x) & = & {{M_o^2} \over {2{H_o}^2}} (2x + e^{-2x} -1),  \\
and\qquad \lambda & = & {{G{M_o}^2} \over {6 \pi}}.   
\eeq
In deriving (55) it is assumed that ${M_o}^2 << {H_o}^2$ and 
$ terms \sim {({M_o \over H_o})}^4$ are neglected.  Assuming $\vert x \vert < 1$
and substituting $Q = {q \over {\sqrt{\lambda}}}$, we find using (55) and (57)
\be
\left[ {{\partial^2} \over {\partial Q^2}} - {1 \over Q^2}{{\partial^2}
\over {\partial x^2}} - Q^2 (1 - Q^2 U(x)) \right] \Psi = 0
\ee
where
\be
U(x) = \lambda (1 - m^2 x^2)
\ee
and
\be
m^2 = {{{M_o}^2} \over {{H_o}^2}}.
\ee
As discussed in the previous section, we convert (55) in the Schroedinger form 
substituting the WKB form
\be
\Psi = \exp {(i {\sum_{n=o}^{\infty}} M^{1-n} A_n)}\psi(Q,x).
\ee
Collecting terms in different order in $M$ we obtain for 
${M^2} - order \, {{\partial A_o} \over {\partial x}} = 0$, which implies 
that $A_o$ is purely a functional of the gravitational field and ${M^1} -\, order$
gives the source free Hamilton-Jacobi equation
\be
{({{\partial A_o} \over {\partial Q}})^2} + Q^2 = 0.
\ee
We introduce a time operator as
\be
{d \over {dt}} = - {1 \over Q}{{\partial A_o} \over {\partial Q}}
{{\partial} \over {\partial Q}}
\ee
and substitute (63) in (60) to get 
 
\be
i{{\partial \psi(Q,x)} \over {\partial t}} = \left[ {-{1 \over {2Q^3}} 
{{\partial^2} \over {\partial x^2}}} + {Q^3 \over 2}U(x) + {1 \over {2Q}} 
\right] \psi (Q,x).
\ee
In obtaining (66) we have neglected 
$terms \, \sim {{{\partial^2} \psi} \over \partial Q^2} $. After the WKB
reduction we have set $M = 1$. Thus we have 
\be
\Psi = \psi_o (Q) \psi (Q,x)
\ee
where
\be
\psi_o = e^{i A_o (Q)}.
\ee
From (64) and (68) one finds 
\be
\psi_o (Q) = e^{\pm {Q^2 \over 2}}.
\ee
In terms of the $q$ variable
\be
\psi_o (q) = e^{\pm {q^2 \over {2 \lambda}}}.
\ee
The solution with the negative sign in the exponent was obtained by Vilenkin 
with the assumption that $q << 1$ and 
\be
{{\partial \Psi} \over {\partial x}} (0,x) = 0
\ee
and the tunneling solution exponentially decreases in the classically forbidden region.
From (55) the classically forbidden region lies between 
$q = 0 \,\; and\; \, q \simeq 1 + {1 \over 2} \mu^2 (x)$  and acts as turning points.
\par
Now to solve (66) we make a Gaussian ansatz
\be
\psi = N (t) e^{- {{\Omega (t)} \over 2} x^2}.
\ee
Substituting this in (66) leads to coupled equations for $\Omega$ and $N$
\be
i{d \over {dt}} \ln N = {\Omega \over {Q^3}} + {Q^3 \lambda}  
+ {1 \over Q}
\ee
and
\be
i {\dot{\Omega}} = {{\Omega^2 + Q^6 \lambda m^2} \over Q^3}.
\ee
In (73) and (74), $t$ is the WKB time defined by (65) and parameterizes the classical 
trajectories in the minisuperspace which is now spanned by $Q$. With 
the ansatz
\be
\Omega = - i Q^3 {{\dot{y}} \over y}
\ee
one finds from (74)
\be
{\ddot{y}} + 3 {{\dot{Q}} \over Q}{{\dot{y}}} - \lambda m^2 y = 0.
\ee
Introducing the conformal time $\eta$ by the relation $dt = Q d\eta$, we get from 
(76)
\be
y'' + 2 {{Q'} \over Q} y' - \lambda m^2 Q^2 y = 0
\ee
where prime denotes a derivative with respect to $\eta$. We now specify the 
model by specifying
$Q = - {1 \over {\sqrt{\lambda}\; \eta}}$ such that the background undergoes an 
exponential expansion. Equation (76) now reduces to
\be
y'' - {2 \over \eta} y' - {m^2 \over \eta^2} y = 0
\ee
and is solved by
\be
y = \eta^{{3/2} \pm {\sqrt{{9/4} + m^2}}}.
\ee
Approximating $\sqrt{{9/4} + m^2} = ({3/2} +{{m^2}/3}) 
\,(since \;\, m^2 = {{{M_o}^2} \over {{H_o}^2}} << 1)$, and using (75) in conformal time
coordinate we get,
\be
\Omega = -i m^2 \sqrt{\lambda} {Q^3 \over 3}.
\ee
As $\Omega$ is imaginary the state (72) will not be normalizable. One of the ways to 
obtain the real part in $\Omega$ is to consider various mode solutions of the scalar 
field $x$ as in the work of Kiefer [3]. However we will follow a different line  
based on our previous work [29] having interpretational significance along with 
settling the boundary conditions. Remembering that $m^2 \sqrt{\lambda}$ is a very
small quantity, we approximate (73) for large $Q$ 
\be
i {d \over {dt}} \ln N = {\Omega^3 \lambda} .
\ee
With ${d \over {dt}} = \sqrt{\lambda} Q {d \over {dQ}}$, one finds from (81)
\be
N = N_o \exp { \left[ - {{i Q^3 \sqrt{\lambda}} \over 3} \right]}
\ee
so that
\be
\psi = N_o \exp { \left[- {{i Q^3 \sqrt{\lambda}} \over 3} 
+ {{i m^2 \sqrt{\lambda}Q^3} \over 6}x^2 \right] }
\ee
where $N_o$ is a constant to be determined. The bracketed term in the 
exponent (83) is
approximated as 
\beq
S_{eff} & = & - {{i Q^3 \sqrt{\lambda}} \over 3} 
+ {{i m^2 \sqrt{\lambda}Q^3} \over 6}x^2 \nonumber \\
& = & -{i \over 3} {\left[ {Q^2 \lambda}(1 - \mu^2 (x)) \right]}^{3/2} 
{{1 + \mu^2} \over \lambda} \nonumber \\
& \simeq & -{i \over 3} {\left[ {Q^2 \lambda}(1 - \mu^2 ) - 1 \right]}^{3/2} 
{{1 + \mu^2 (x)} \over \lambda} 
\eeq
In our previous work [29] we have shown that the multiple reflections between the
turning points will contribute to the normalization factor  and are supposed to 
arise from the wormhole contributions. Using (84) and the result of [29] 
the normalization constant $N_o$, according to wormhole dominance proposal,
turns out to be 
\be
N_o = {{\exp {S_{eff} (Q_x,0)}} \over 
{1 - \exp {\left[ 2 S_{eff} (Q,0) \right]}} }
\ee
where
\be
S_{eff} (Q_x, 0) = S_{eff}(Q){\vert_{0}^{Q_x}} 
\ee
and $Q = 0$ and $Q = Q_x = {1 \over {(\lambda (1 - \mu^2 (x)))}^{1 \over 2}}$
are the turning points as can be seen from (55) and (57). We thus get from (85)
and (83)
\be
\psi \sim \exp { \{ +{1 \over {3 \lambda}} (1 + \mu^2 (x)) \left[ 1 - 
i (q - q^2 \mu^2 -1)^{3/2} \right] \}}
\ee
For simplicity we have dropped the denominator of (85) but is 
nonetheless important while discussing the normalization. Eqn.(87) continued in
classically forbidden region gives
\be
\psi \sim {\exp{\{{1 \over {3\lambda}}(1+\mu^2(x))\left[1-{(1-q^2+q^2\mu^2)}^{3/2}\right]\}}
\over {(1- \exp{\{{2/{3\lambda}}}(1+\mu^2(x))\})}}
\ee
Eqn.(87) and (88) are the wavefunctions that one gets from the Hawking's proposal.
\par
We can now discuss the curvature fluctuations in the newly-born universe. From
(87), the probability of nucleation with a certain value of $x$ is proportional 
to 
\be
{\vert \psi \vert}^2 \propto \exp { \left[ -{{2 \mu^2 (x)} \over {3 \lambda}}
\right]}.
\ee
In obtaining (89), a factor $({e^{-{2\over {3\lambda}}(1+\mu^2)} - 1})^2$ in the
denominator is kept as multiplying (89).
For $x << 1, \mu^2 (x) = m^2 x^2 = {({{M_O}^2 \over {H_o}^2}) x^2}$ and we get
\be
{\vert \psi \vert}^2 \propto \exp { \left[ -{{4 \pi x^2} \over {G{H_o}^2}}
\right]}.
\ee
Now it follows from (56) that $x = {1 \over 2}{{\delta R} \over {R_o}}$, where 
$\delta R = R - R_o$ is the curvature fluctuations. Hence we can write
\be
< {({{\delta R} \over {R_o}})}^2> = 4 <x^2> = {{G{H_o}^2} \over {2
\pi}}.
\ee
Exactly this result was obtained by Vilenkin from the tunneling proposal, but
our wavefunction corresponds to the Hawking's proposal.
\par
Though the result (91) is identical with the Vilenkin, there are some marked 
interesting  differences. In Vilenkin [7] ${1 \over {3\lambda}} (1 + \mu^2 (x))$
term is added in (87) to get the behaviour $S \rightarrow {{q^2} \over {2\lambda}}$
for $q \rightarrow 0$.
In our approach when the denominator in (87) is taken into account, 
$S\rightarrow-{q^2 \over {2\lambda}}\; as \; q\rightarrow 0$.
If we set aside the denominator in (87), the wavefunction looks apparently
as the Vilenkin wavefunction. By analytic continuation, in the classically allowed
region, Vilenkin gets back the term $-{1 \over {3\lambda}}(1+\mu^2)$ in the exponent
to find   
\be
\psi \sim \exp { \{ -{1 \over {3 \lambda}} (1 + \mu^2 (x)) \left[ 1 -i 
(q^2 - q^2 \mu^2 -1)^{3/2} \right] \}}.
\ee
This analytic continuation seems questionable. 
Thus we see that the Gaussian ansatz and the normalization prescription serve as 
the boundary conditions to settle between the two current proposals in favour of
the Hawking's prescription. We also stressed this in our previous work [29]. The
present work works out explicitly the equivalence between the timeless
Wheeler-DeWitt equation (55) and the time contained Wheeler-DeWitt equation (66)
and hence the time parameter prescription as mentioned in this work suits also 
in the frame work of the quantum cosmology.
\section{\bf{Discussion}}
We find that the normalization factor arises from the repeated reflections between 
the turning points and hence corresponds to the higher order corrections related to
the quantum fluctuations in the WKB description. In our previous work [29] we have shown
that the factor $N_o$ can be interpreted as contributions from the wormholes using 
Klebanov and Coleman's arguments [30,31]. Apart from the wormhole picture, it can be said 
that the quantum force has its origin in the curvature fluctuations with `to and fro'
motion within $0 < q < 1 + {1 \over 2}\mu^2 (x)$ and it necessitates to uphold the 
probabilistic interpretation. Leaving aside the interpretational hindrance,
`to and fro motion' is assigned to terms like $e^{iS}$ and $e^{-iS}$ but the 
Gaussian ansatz required only the form (72) in the classically allowed region
as if there is a suppression of interference terms. It has been argued that 
some boundary conditions at small scale [32] would lead to the quantum effects
in the vicinity of the turning points. Thus the only way to interpret $e^{iS}$
and $e^{-iS}$ superposition [i.e., the quantum effects] is to turn back towards
Klebanov and Coleman's arguments of wormholes connections around $q \approx 0$.
There is also an objection about the `to and fro' motion with respect to time 
because there is no external time parameter with respect to which the universe
can turn. But our calculation in the time variable description correctly obtains 
the wavefunction both in the classically allowed and in the forbidden region. 
The classical turnings points serve as a clue to obtain the peculiar behaviour 
of wave packets at a late time $(q >> 1)$ whereas the quantum turning point 
(we identify it to be $q = 0)$ leads to `to and fro' quantum fluctuations. 
The Euclidean and the Lorentzian time both are manifestation of the spacetime 
structure latent
in the description of the Wheeler-DeWitt equation. There is also a question about 
the influence of excited matter states and whether they lead to decoherence.
In the present work we prove the decoherence in reverse way i.e., leading to the 
form (88) 
from the Gaussian ansatz (72); however we have been able to show [33]
that the presence of effective decoherence even in the Starobinsky
description, using the technique of ref.[16].
Perhaps the arrow of time starts functioning 
from the classical turning point when the universe emerges from the Euclidean to 
the Lorentzian regime with a signature change.
It should be pointed out that the denominator in (88) arises from the multiple 
reflections from the turning points and indicates decisively in favour of the
Hawking wavefunctions. If we accept the Klebanov and Coleman's [30,31] arguments,
it would not be unjustified to comment that wormholes contribution
generates
the initial randomness, and the quantum behaviour thus generated even persists
in the classical spacetime. The emergence of the classical universe is then
explained through the decoherence to suppress the interference of 
$\exp{iS}\; and\; \exp{-iS}$ like terms. The code of the decoherence is seeded
in the normalization constant rather than in the initial conditions.
\par
It is worthwhile to point out that, at the semiclassical level, the Starobinsky
inflationary scenario has been criticized [34] on the ground that the inflationary
solutions are not perturbatively expandable in the parameter of ``quantum 
corrections'' terms in the equation of motion or the Lagrangian. There are also some 
questions of treating the higher order terms in the Lagrangian on an equal 
footing with the Einstein terms. Even then, studying the quantum cosmology of the 
Starobinsky description from the standpoint of quantum to classical transition
would help us rethink about the criticism labelled against the model. Leaving 
aside this fact, it is instructive to look at the model as a toy example, to 
understand the current boundary condition proposals, decoherence mechanism and 
also the origin of the quantum force in the early universe.
\par                                                                 
Further there are also some drawbacks in expanding the Wheeler-DeWitt 
wavefunction as a power of Planck mass as in (43) and in obtaining the 
Schroedinger-Wheeler-DeWitt equation (SWD) 
from it. Recently [26,35] we have been able to obtain the SWD equation 
without using the WD equation and also the expansion  (43) and it has been found 
that the SWD wavefunction can be defined on the standard Hilbert space of quantum 
mechanics. Though the Starobinsky description itself has some drawbacks as 
discussed in [30], our attempt is to understand the quantum to classical 
transition in the framework of the model and to investigate which of the boundary 
conditions proposal suits to the inflationary description. 
We find that (i) the classical Starobinsky model is consistent with the 
quantum description, (ii) the wormhole dominance proposal [29] correctly 
connects the Wheeler-DeWitt wavefunction and the Schroedinger-Wheeler-DeWitt 
wavefunction  in the respective regime, (iii) the curvature fluctuation is 
correctly reproduced from the wormhole dominance wavefunction, (iv) the 
wormholes initiate the quantum randomness in the initial stage (i.e., 
$0<a<H^{-1}$) and (v) the decoherence is effectively [33] reproduced in 
the Starobinsky description, provided we accept the wormhole dominance 
proposal keeping intact the fruits of the inflationary scenario.  
\section{\bf{Acknowledgment}}
Dr. S. Biswas is grateful to Prof. P. Dasgupta for stimulating discussions.
A. Shaw acknowledges the financial support from ICSC World Laboratory,
LAUSSANE, Switzerland during the course of the work.
\newpage
\begin{center}
{\bf{References}}
\end{center}
{\obeylines\tt\obeyspaces 
1. A.H. Guth, Phys. Rev.{\bf{D23}}, 347(1981)
2. A.A. Starobinsky, Phys. Lett.{\bf{B91}}, 99(1980)
3. W. Boucher, G.W. Gibbons and G.T. Horewitz, Phys. Rev.{\bf{D30}}, 
   \qquad 2447(1984)
4. M.B. Mijic, M.S. Morris and W.M. Suen, Phys. Rev.{\bf{D34}},
   \qquad 2934(1986) 
5. K. Maeda, Phys. Rev.{\bf{D37}}, 858(1988)
6. A. Vilenkin, Phys. Rev.{\bf{D27}}, 2848(1983)
7. A. Vilenkin, Phys. Rev. {\bf{D32}}, 2511(1985) 
8. A. Vilenkin, Phys. Rev. {\bf{D37}}, 888(1987)
9. S.W. Hawking, Nucl. Phys. {\bf{B239}}, 257(1984); 
   \qquad S.W. Hawking and D.N. Page, Nucl. Phys.{\bf{B264}}, 184(1986)
10. A.A. Starobinsky and H.J. Schmidt, Class. Quantum. Grav. 
    \qquad {\bf{4}}, 694(1987)
11. R.M. Wald, Phys. Rev.{\bf{D28}}, 2118(1983) 
12. E. Mottola, Phys. Rev.{\bf{D31}}, 754(1985)
13. D.E. Nevelli, Phys. Rev.{\bf{D8}}, 1695(1984)
14. J.J. Halliwell and S.W. Hawking, Phys. Rev.{\bf{D31}}, 1777(1985)
15. C. Kiefer, Class. Quantum. Grav.{\bf{4}}, 1369(1987)
16. C. Kiefer, Phys. Rev.{\bf{D46}} (1992), 1658; {\bf{D45}}, 2044(1992)
17. C. Kiefer, in `Time, Temporality, Now' edited H Atmanspacher 
    \qquad and E Ruhnau (Springer, Berlin), pp 227-240 (1997)
18. C. Kiefer, D. Polarski and A.A. Starobinsky, gr-qc/9802003
19. S. Wada, Nucl. Phys. {\bf{B276}}, 729(1986) 
20. K.V. Kuchar, in:Proceeding of the fourth Canadian Conference 
    \qquad on General Relativity and Relativity Astrophysics, 
    \qquad ed. by Kunstatter G Vincent D Williams J 
    \qquad (World Scientific, Singapore), p 211-314 (1992)
21. E. Santamato, Phys. Rev.{\bf{D29}}, 216(1984)
22. E. Madelung, Z. Phys.{\bf{40}}, 332(1926)
23. D. Bohm, Phys. Rev.{\bf{85}}, 166(1952)
24. I. Feynes, Z. Phys.{\bf{132}}, 81(1952) 
    \qquad D. Kershaw, Phys. Rev.{\bf{B136}}, 1850(1962) 
25. E. Nelson, Phys. Rev.{\bf{150}}, 1079(1966)
26. S. Biswas, A. Shaw, B. Modak and D. Biswas, ``Quantum Gravity 
    \qquad Equation in Schroedinger Form in Minisuperspace Description''
    \qquad gr-qc/9906011
27. J. Butterfield and C.J. Isham, gr-qc/9901024  
28. C. J. Isham, gr-qc/9210011 
29. S. Biswas, B. Modak and D. Biswas, Phys. Rev.{\bf{D55}},
    \qquad 4673(1996)
30. I. Klebanov, L. Susskind and T. Banks, Nucl. Phys.{\bf{B317}}, 
    \qquad 665(1989)
31. S. Coleman, Nucl. Phys.{\bf{B310}}, 643(1988)
32. H.D. Conradi and H.D. Zeh, Phys. Lett.{\bf{A154}}, 321(1991)
33. S. Biswas, A. Shaw and B. Modak, `Decoherence in Starobinsky Model'
    \qquad General Relativity and Gravitation (in press) (Vol. 31, 1999)
34. J.Z. Simon, Phys. Rev.{\bf{D45}}, 1953(1992);
    \qquad L. Parker and J.Z. Simon, Phys. Rev. {\bf{D47}}, 1339(1993)
35. S. Biswas, A. Shaw and B. Modak, ``Time in Quantum Gravity'', 
    \qquad gr-qc/9906010.}
\end{document}